# Switching Transients in Constrained Transformer-Line/Cable Configurations

**Y. XIANG**         **L. WU   K. VELITSIKAKIS**         **A.L.J. JANSSEN**
TU Eindhoven              DNV GL                              Liander
                       the Netherlands

**SUMMARY**

This paper investigates the transient phenomena that occur in two special cases in the Netherlands: (A) during the energization of a power transformer via a cable feeder and (B) the energization of a power transformer together with an overhead line (OHL). In Case A a 7 km long 150 kV cable and a 150/50 kV transformer are connected and energized at the same time. In Case B a 150/50 kV transformer and a short 50 kV OHL are connected and energized simultaneously. The reason behind this kind of situations is related to space restrictions and cost efficiency.

Focus is given to the theoretical explanation of the high frequency and resonance phenomena during different stages of switching transients for both cases. Proper EMT models are then built to replicate the real network configurations within the Dutch system. Different scenarios and configurations are simulated, and the transient overvoltages are observed. It could be concluded that resonance phenomena occur during the energization of the combinations cable-transformer and transformer-OHL, due to the interaction between the transformer inductance and the capacitance of the cable or OHL. Moreover, it is noted that the tap position of the transformer has a large impact on the magnitude of the overvoltage at the secondary side of the transformer. In general, a higher tap position leads to a higher switching overvoltage, just by the ratio of the windings. Furthermore, the tap position has a slight influence on the leakage inductance of the transformer and a moderate influence on the frequency dependent transfer function.

Energizing the transformer with a long cable between circuit breaker and transformer gives a doubling of the voltage step-function by the reflection at the transformer terminals. Compared to a situation without cable, this leads to a considerable increase of the stresses both in the transformer and to the external equipment, but it is still far below the specified design values. The connections at the secondary side tend to reduce the initial overvoltages, as they decrease and damp the capacitively transferred voltage step function. But the inherent 1-cos function from the natural frequency (cable/line capacitance and leakage inductance) causes an overswing that leads to an increase in the switching overvoltages. Moreover, resonance between the connections (primary and/or secondary side) and the transformer Eigen-frequencies may lead to severe switching overvoltages, but that did not appear in the cases described.

anton.janssen@alliander.com

# 1 Introduction

The energizing process of a power transformer is a regular and planned switching operation. More specifically, without special pre-cautions, during its energization, a stand-alone power transformer withdraws high inrush currents that may cause a temporary voltage dip in the system as well as slow front and temporary overvoltages. The inrush current and the voltage profile are dependent on the equivalent network impedance, the saturated transformer magnetizing inductance, the residual magnetic flux in the transformer core, switching moment, pole asymmetry of the core and neutral treatment, the pole discrepancy of the circuit breaker and the loading conditions of the transformer. Furthermore, due to the sympathetic interaction, the large voltage dip might lead to the saturation of transformers in parallel or in series to the one that is energized.

More and more due to space restrictions and/or problems with permits, and also because of economic reasons or forced by time pressure, rather simple methods to connect transformers to the existing transmission systems are emerging. For instance, transformers may be directly connected to a nearby overhead line (OHL) or an underground cable (UGC), albeit by means of a circuit breaker. In addition, conditions where the transformer is connected by means of a disconnector can also be found. Such configurations may lead to special switching transients, which arise due to the interaction between the transformer and OHL/UGC at the primary or secondary side.

Previous study [1] investigated this problem and compared the simulation results with real measurement data. However, the scenarios considered are limited. In this paper EMT studies are performed to investigate these transient phenomena, by properly simulating the parts of the Dutch transmission and distribution system under study. The EMT software EMTP/ATP [2] is used throughout the studies, where the theoretical background of the transient analysis has been presented in [3-11], although the information about the transformer transient behaviour is limited to some Frequency Response Analysis (FRA) measurements in the past, reported by the authors of [12]. Another publication on a similar switching condition can be found in [13].

# 2 Two Special Cases in the Dutch system

Within the Dutch distribution system operator Liander, special transformer switching transient phenomena have been found in two cases, as shown in Figure 1. In case A the 150-kV primary side of a transformer is directly connected to a 7.1 km long UGC, with the circuit breaker at the far end. By closing the circuit breaker, the transformer is energized together with the UGC feeding the transformer. There exist two systems in parallel, which normally are energized, feeding a 50 kV grid.

In case B the 50-kV secondary side of another transformer is directly connected to a 13 km long OHL, without circuit breaker installed in between. The circuit breaker is installed at the 150-kV primary side of the transformer. By closing that circuit breaker, the transformer is energized together with the OHL connected to the secondary side. In both cases, the neutral at the 50 kV side is floating. Testing and simulations are based on another situation with a 4 km long OHL, where a circuit breaker is available for comparison. Here both circuits of the 50 kV line are connected in parallel, since nowadays there is only one 150/50 kV transformer available.

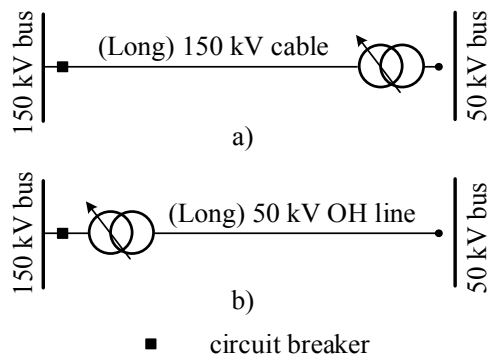

*Figure 1 Two cases where a transformer is directly connected to an UGC or OHL*

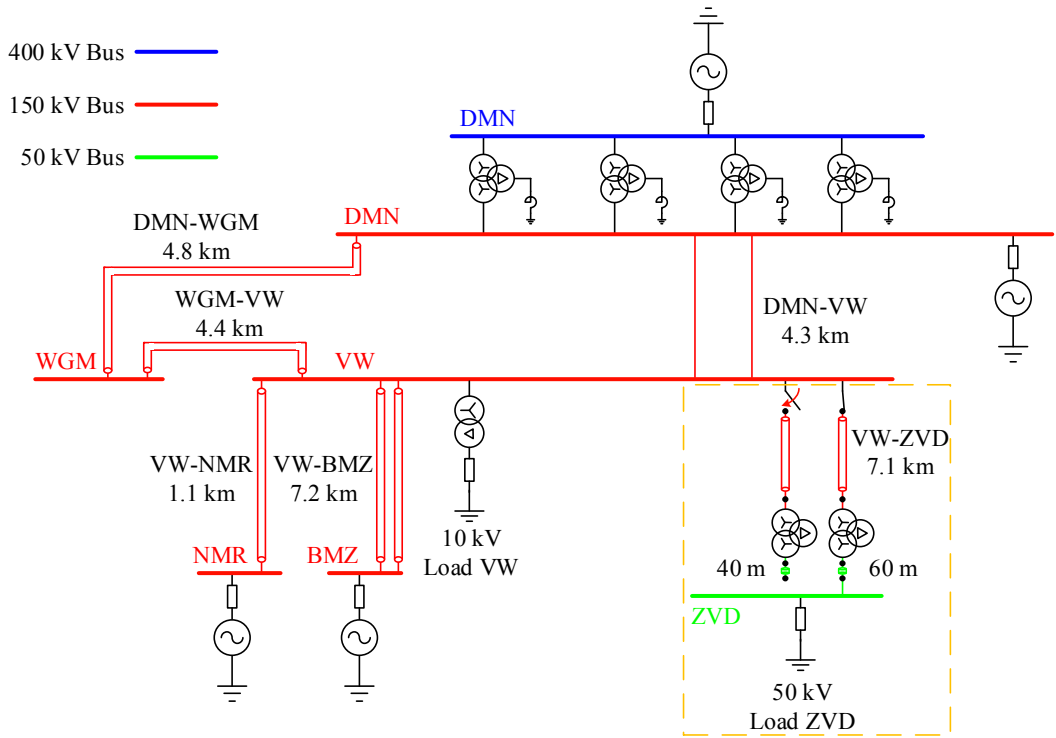

*Figure 2 Simplified network model for switching activity between VW and ZVD (Case A), with investigated connection in orange dashed box*

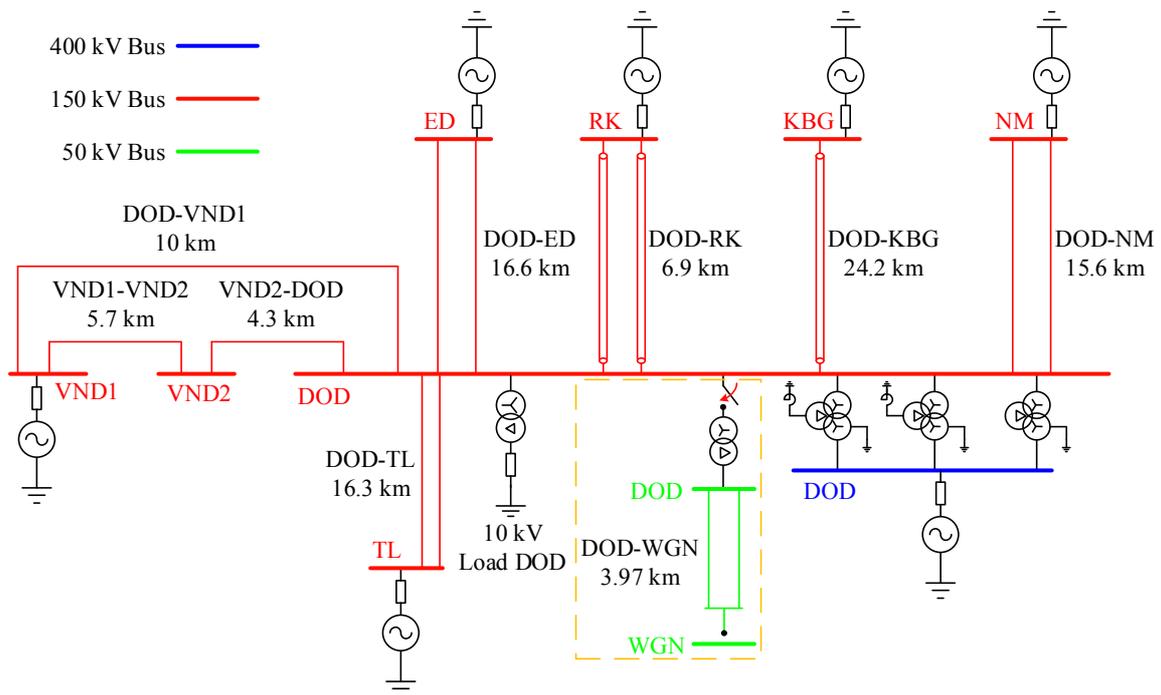

*Figure 3 Simplified network model for switching activity between DOD and WGN (Case B), with investigated connection in orange dashed box*

Table 1 Parameters of the transformer in VW-ZVD

|  | ZVD | DDW |  | ZVD | DDW |
|---|---|---|---|---|---|
| Rated Power (MVA) | 100/100/30 | 80 | No load losses (kW) | 38.2 | 71 |
| Voltage (kV), tap 11 | 150/52.5/11 | 150/52.5 | Leakage impedance (HV-LV, %) | 14.19 | 13.8 |
| tap 1 (kV) | 172.5 | 172.5 | Leakage impedance tap 1 (%) | 15.19 | x |
| tap 21 (kV) | 127.5 | 127.5 | Leakage impedance tap 21 (%) | 13.43 | x |
| Magnetizing Current (%) | 0.3 | 0.3 | Short-circuit losses (HV-LV, kW) | 267.4 | 287 |

Slow front overvoltages and resonances have been feared during switching transients in both cases. It has to be noticed that a proper adjustment of the tap changer was suggested in the past, as a possible action to (partially) mitigate overvoltages, when necessary for Case A.

Figure 2 shows the actual network model that was developed for Case A, with the studied connection VW-ZVD (to be switched in) in orange dashed box. In a normal switching condition, one cable-transformer connection will be in service, while the other is energized (for instance after maintenance). If the second connection is put into service, transformer and cable will be energized together. The short-circuit current level at VW substation is about 33 kA (rms). Table 1 gives the transformer data.

Figure 3 shows the actual network model that was developed for Case B, with the studied connection DOD-WGN in orange dashed box. At the 150-kV substation DOD, the transformer secondary side is directly connected to the 50-kV double-circuit overhead line to the 50-kV substation WGN; the transformer and the OHL will be energized together. The short-circuit current level at substation DOD is 42 kA (rms). The transformer data are listed in Table 1.

## 3 Theoretical Analysis
### 3.1 Transformer without cable or OH-line at secondary side

Energizing an unloaded transformer at the HV-side is a normal and regular switching procedure. The 150/50 kV transformer energization by the 150 kV circuit breaker is taken as a reference. Pre-arcing of the circuit breaker contacts generates a voltage step function that will be transferred to the 50 kV-side by the transformer's transfer function. During the very fast initial stage of the switching process the transfer function is dominated by the transformer's distributed capacitances [14]. At high frequencies, the most important capacitances are the mutual capacitance between the windings and the surge capacitance, as can be measured from the 50 kV side. In the MHz range, FRA measurements show a ratio (transfer function) of 0.2, which means that 20% of the step voltage applied at the primary side will initially appear at the secondary side. In this case, it is less than the ratio of the winding turns, that determines the (inductive) transfer function for the slower phenomena.

A transformer consists of numerous self and mutual inductances as well as capacitances between the windings and the coils of different voltages, different phases and other parts (tank, core, leads, etc.). All the possible combinations of inductances and capacitances give Eigen-frequencies, with a few dominant resonances. For example: the parallel resonance of the main inductance (short-circuit inductance or open circuit inductance) with the total (or surge) capacitance, that forms a high impedance at the resonance frequency. In other cases, some anti-resonance frequencies give a low impedance and are represented by the series connection of an inductance with a capacitance and a resistance. For switching voltages at the transformer terminals, frequencies between some kHz up to one hundred kHz are usually of interest. For a limited frequency range, the admittances $Y(\omega)$ from FRA measurements, are shown in Figure 4.

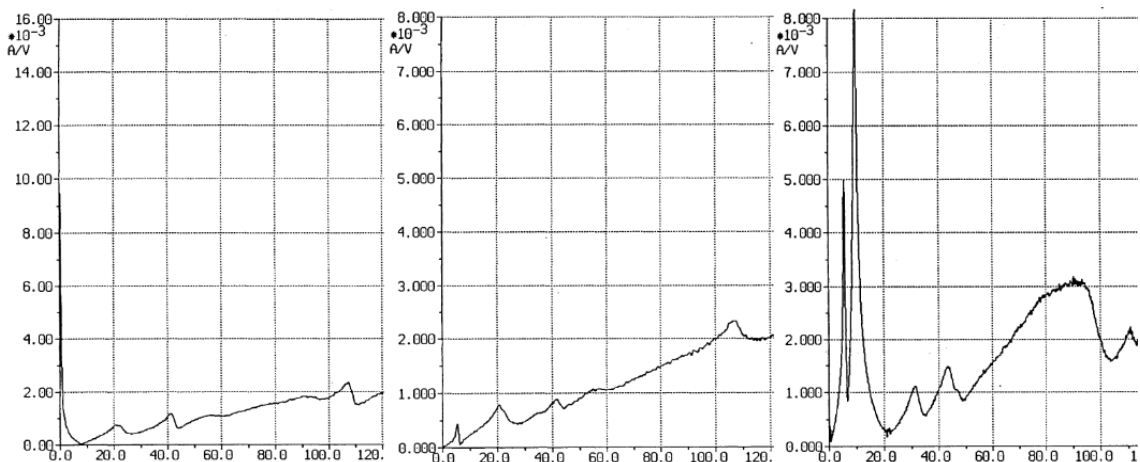

*Figure 4 Admittance in $10^{-3}$ A/V as function of the frequency in kHz, tap 11, unloaded transformer measured at 150 kV side (left) and 50 kV side (right) and short-circuited measured at 150 kV side (middle)*

More important is the transfer function H(ω), where peak values coincide with minima in the admittance measurements. The transfer function is given for three different tap positions in Figure 5.

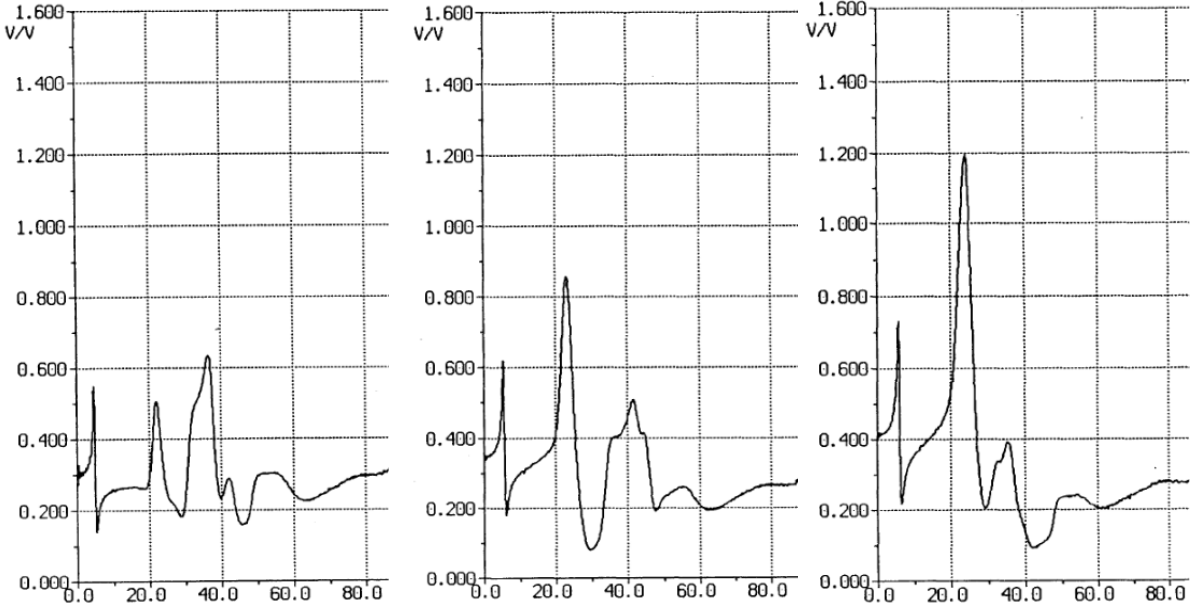

*Figure 5 Transfer function from 150 to 50 kV as function of the frequency in kHz for the tap positions: 1, 11, 21*

Excitation with a certain resonance frequency leads to a response at the unloaded 50 kV side higher than corresponding to the power frequency ratio of the windings (visible at 50 Hz). At a higher tap position this response is even more extreme (up to a factor 3). Thus, the tap position plays a role with respect to the power frequency ratio (a higher tap gives more severe voltage stresses at 50 kV-side), with respect to the transformer short-circuit impedance (higher tap positions give slightly lower inductances) and with respect to the frequency dependent impedance and transfer function.

By modelling the transfer function of the highest tap in the frequency range illustrated in Figure 5 (right), the most severe voltage stress due to a step function can be simulated. Figure 6a shows a simulation of the response to a step function for the first 120 μs, including the ratio 0.2 for the very high frequencies, but excluding the (limited) influence of the transfer function in the range above 60 kHz. The transformer's main Eigen-frequency is visible and peak values are up to the winding ratio (which is 0.412). However, with a relatively long cable between circuit breaker and transformer, the travelling waves will reflect at the transformer terminals, thus causing a 2.0 p.u. step function. Figure 6b gives the response at the 50 kV side, when taking 1 p.u.=√(⅔)*150 kV as a reference. Note that no damping of the travelling waves is considered. The peak value in Figure 6b (at 50 kV side) is 0.82*122 kV$_p$ = 100 kV$_p$. When taking the damping into consideration and knowing from figure 9b, that the initial voltage step is equal to 185 kV$_p$ (1.5 p.u.), the peak value at 50 kV will be less than 75 kV$_p$. It has to be compared with 80% of the lightning impulse withstand value of 325 kV$_p$.

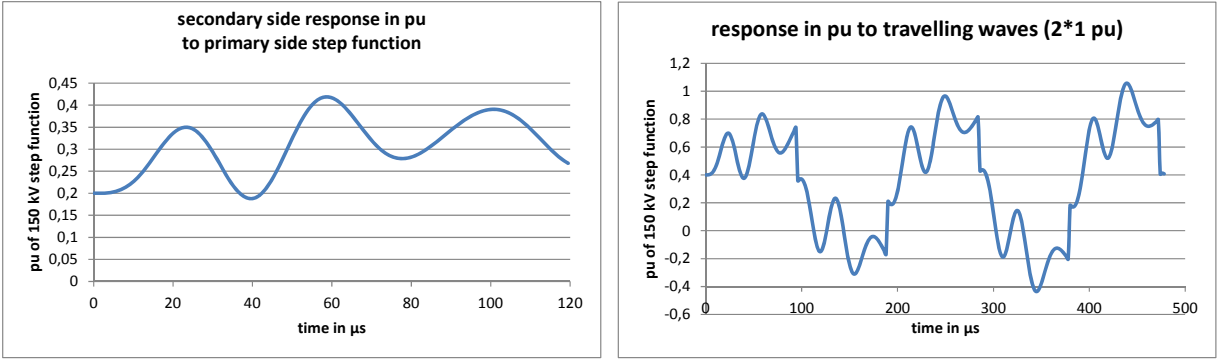

*Figure 6 a) Response at 50 kV side to 1 p.u. step at 150 kV, b) Response to 2*1 p.u. travelling wave, 2*80 μs*

## 3.2 Transformer with cable or OH-line connected to secondary side

As soon as a cable or an OHL is connected to the secondary side, the transformer Eigen-frequencies are overruled by the interaction between cable or line and the transformer. In order to compare the phenomena, a simulation without the transformer's transient characteristics is given for the middle tap (to be discussed in section 4.1.2). The travelling waves at the 150 kV side are clearly visible at the 50 kV side with an initial peak of 67 kV$_p$.

When the capacitance of a 60 m long XLPE-cable with 1200 mm$^2$ Al (13.2 nF) is connected to the secondary side, the natural frequency caused by the transformer leakage inductance of 12.45 mH (50 kV side) becomes 12.4 kHz. It gives a 1-cos function around 67 kVp, doubling it to a total amplitude of 134 kVp, shown in Figure 8. A shorter length of the 50 kV cable (e.g. 40 m) leads to a higher natural frequency (e.g. 15.2 kHz), but the same amplitude.

In case an OHL is connected to the transformer, the time constant of the line surge impedance and the transformer surge capacitance, which can be deduced from the admittance measurements (Figure 4), will be far less than 1 µs, certainly in the case described with both 50 kV circuits connected in parallel ($Z_{line}$ is approximately 200 Ω). Thus, the transient phenomena will be determined by (i) travelling waves on the OH-line, that is open at the other side, and (ii) the natural frequency of line capacitance (118 nF), and transformer leakage inductance (15 mH), 3.76 kHz, shown in Figure 16 and Figure 17.

The travelling waves are hardly visible, but when zooming in, some deformation of the 1-cos wave-shape can be seen. A step voltage applied to the transformer leakage inductance in series with the line surge impedance causes a 1-exp function with a time constant of L/Z, thus 75 µs. This is far more than twice of the travelling time of the 4 km long OH-line (2*14 µs). The travelling waves, the natural frequency, and their sum are shown in Figure 7.

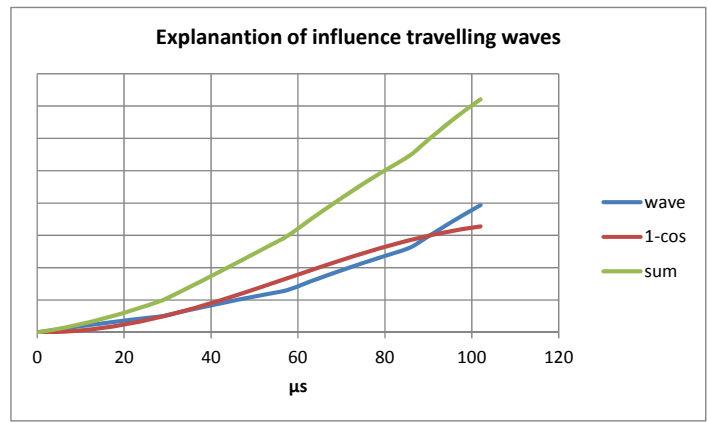

*Figure 7 Explanation on travelling waves, the natural frequency, and their sum*

## 3.3 Line/Cable Model

The Line/Cable Constants routine is used in all line/cable modelling. The detailed tower configuration of the OHLs and layer configuration of UGC are modelled. The commonly used JMarti model is applied for the OHL and Bergeron model, tuned at 5000 Hz, for the 150 kV UGC [2].

## 3.4 Transformer Model

In the simulation study, the BCTRAN model is used [2]. In this model, no capacitive coupling inside the transformer is considered. In the practical cases of this paper, there are cables on both side of the transformer, the cable capacitance is much larger than the transformer stray capacitance. Therefore, the influence of stray capacitance can be reasonably ignored. Moreover, the inductive mutual coupling of the transformer windings on different limbs is also not considered in the simulation. Ferroresonance cannot be simulated in this model.

# 4 Analysis and Simulation Results

## 4.1 VW-ZVD (Case A)

The cable will show a natural frequency caused by the travelling waves, twice forward and backward, with a propagation speed of 150 m/μs, resulting in a cycle time of 190 μs and a frequency of about 5 kHz. This frequency could coincide with one of the transformer's natural frequencies, but this is not the case, as can be learned from Figure 5. The simulations are performed with different lengths of 50 kV cables and different transformer tap positions.

### 4.1.1 Base Case

The simulation in the base case is performed with following settings:
1. Switched object: 150 kV cable, the transformer, and the 60 m 50 kV cable all together.
2. The transformer in parallel is already energized.
3. Tap position of transformer (to be switched in) is at neutral (11).
4. Switching moment at the maximum voltage of phase A.

Figure 8 shows the voltages at the open end of the 50 kV cable. Switching overvoltages up to 3.0 p.u. at the 50 kV side can be encountered. Figure 9 shows the voltages at 150 kV side of the transformer, the reflection of traveling waves can be observed. As shown in Figure 10, at the 150 kV bus in VW a voltage distortion is visible, but that is mainly caused by the capacitance of the switched 150 kV-cable and the short-circuit impedance of the bus in VW, in addition to the travelling waves. There is no difference in the 150 kV voltage at VW between energizing the cable with or without transformer.

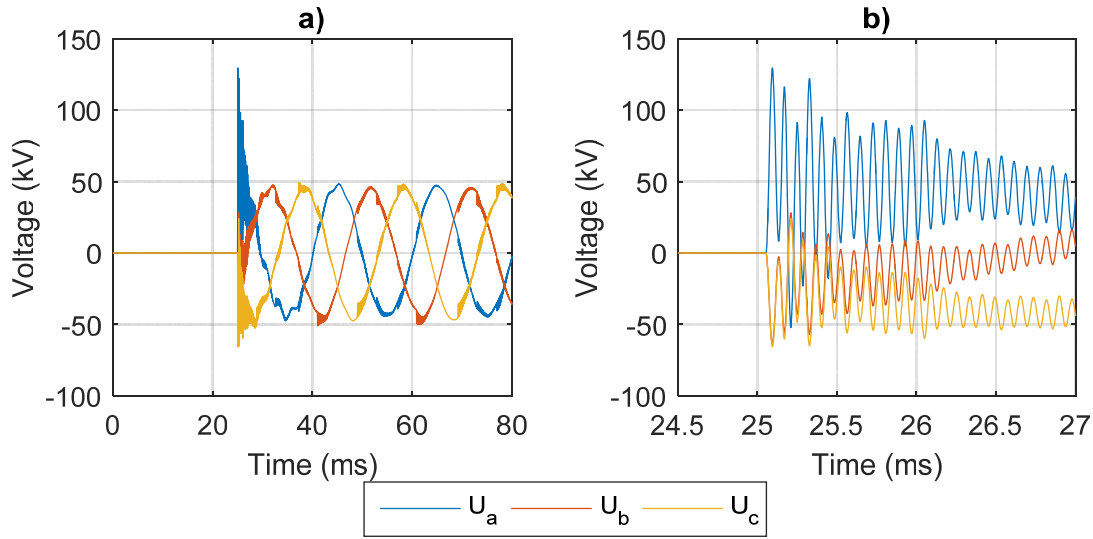

*Figure 8 Voltages at the 50 kV cable for base case: a) total simulation time, b) zoomed in for initial period*

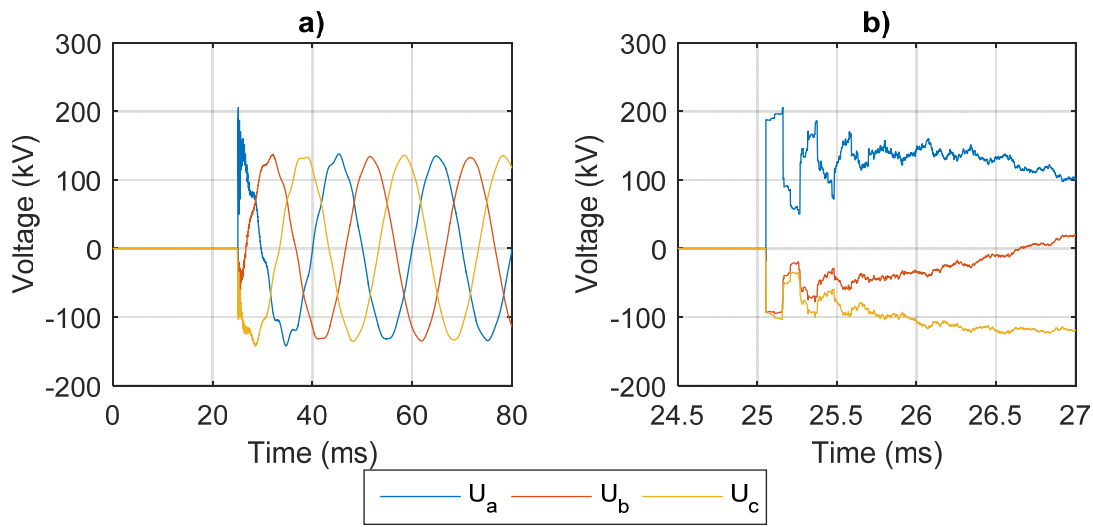

*Figure 9  150 kV side of the transformer for base case: a) total simulation time, b) zoomed in for initial period*

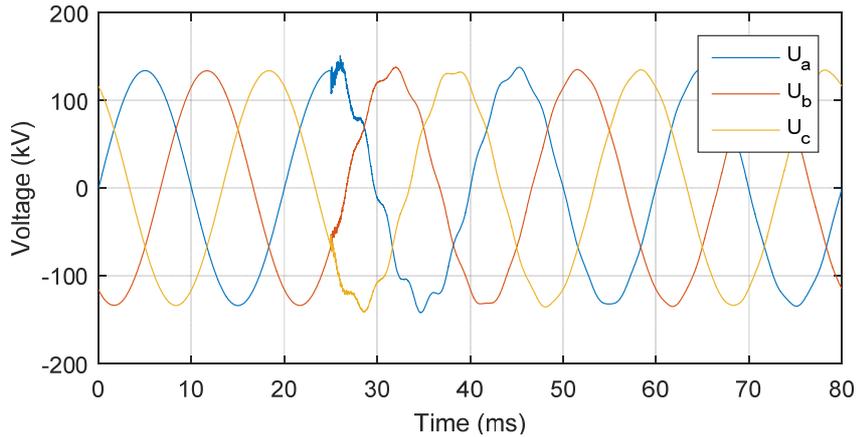

*Figure 10   150 kV busbar at VW for base case: a) total simulation time, b) zoomed in for initial period*

### 4.1.2   Switching the 150 kV Cable and the Transformer without 50 kV Cable

To investigate the influence of the short 60/40 m 50 kV cable on the transient behaviour, a simulation is performed for switching the 150 kV cable and the transformer without 50 kV cable. The phase-to-earth voltages at the open end of the 50 kV side of the transformer are shown in Figure 11. Simulations show a moderate switching overvoltage at the 50 kV side (67 kV$_p$ = 1.5 p.u.), and the waveforms are similar to the travelling waves at the 150 kV side.

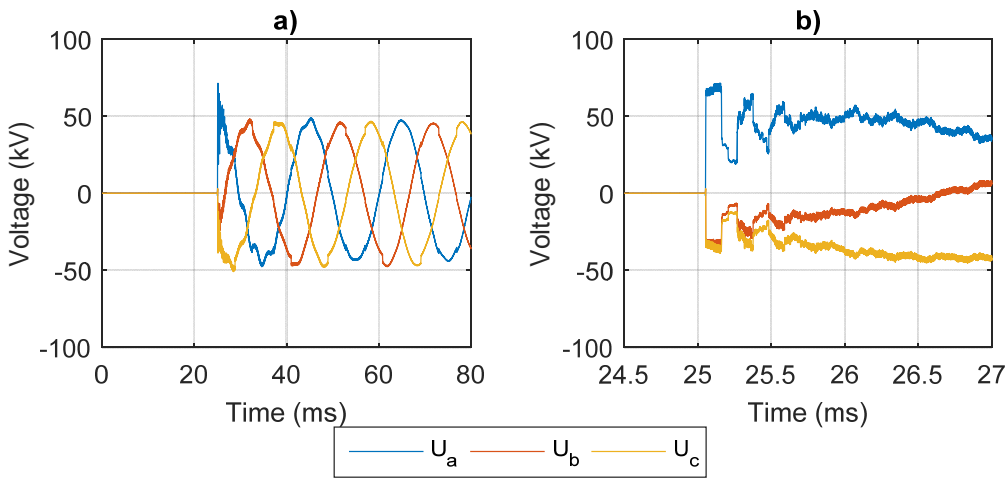

*Figure 11 Voltages at the open end of 50 kV side of the transformer when switching the 150 kV cable and the transformer without 50 kV cable: a) total simulation time, b) zoomed in for initial period*

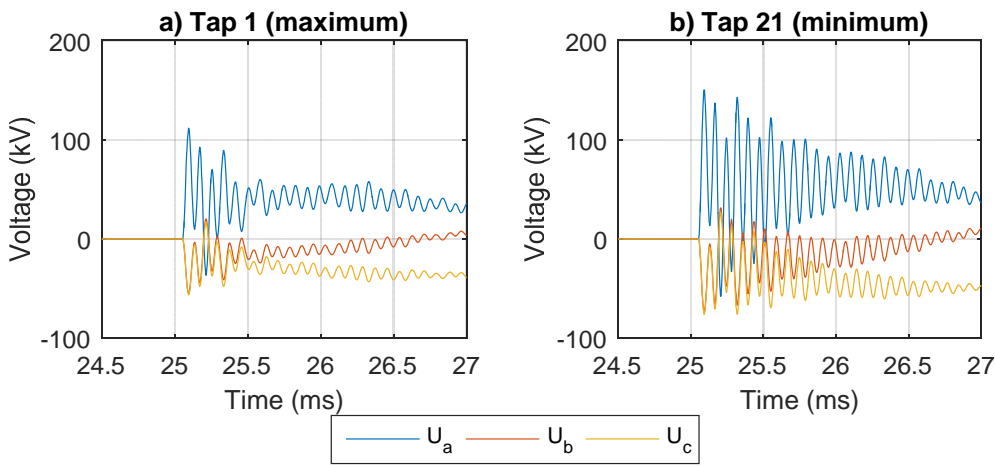

*Figure 12 Voltages at the open end of 50 kV cable for tap positions 1 (a) and 21 (b)*

### 4.1.3 Influence of Tap Positions

Simulations are performed with different tap positions of the transformer. Figure 12 shows voltages at the open end of 50 kV cable when the tap position is at maximum and minimum. It is clear that the overvoltage is less for tap 1(maximum transformer ratio), and that the situation becomes worse for tap 21 (minimum ratio).

## 4.2 DOD-WGN (Case B)

The line length of the open ended double circuit 50 kV OHL (both circuits in parallel) is 4 km and that gives a travel time cycle of 28 µs and a frequency of 35.7 kHz. This frequency could coincide with one of the transformer's Eigen-frequencies. The natural frequency, formed by the leakage inductance and the OHL capacitance, is described before. Combined and separate energization of transformer and OHL have been simulated, as well as variations in the tap positions.

### 4.2.1 Base Case

The simulation in the base case is performed with the following settings:
1. Switched object: the transformer and the 150 kV OHL together.
2. Tap position of transformer is at neutral (11).
3. The switching moment is selected such that the transient overvoltage is at maximum.

Figure 13 shows the voltages at the open end of the 50 kV OHL. The overvoltages can be observed and the oscillation frequency calculated from the Fast Fourier Transformation (FFT) is 3.76 kHz, as calculated before. Similar waveforms can be observed from real measurements, shown in Figure 14. In real situation the damping of the oscillations is faster because the worst case (with minimum damping) is considered in the simulation.

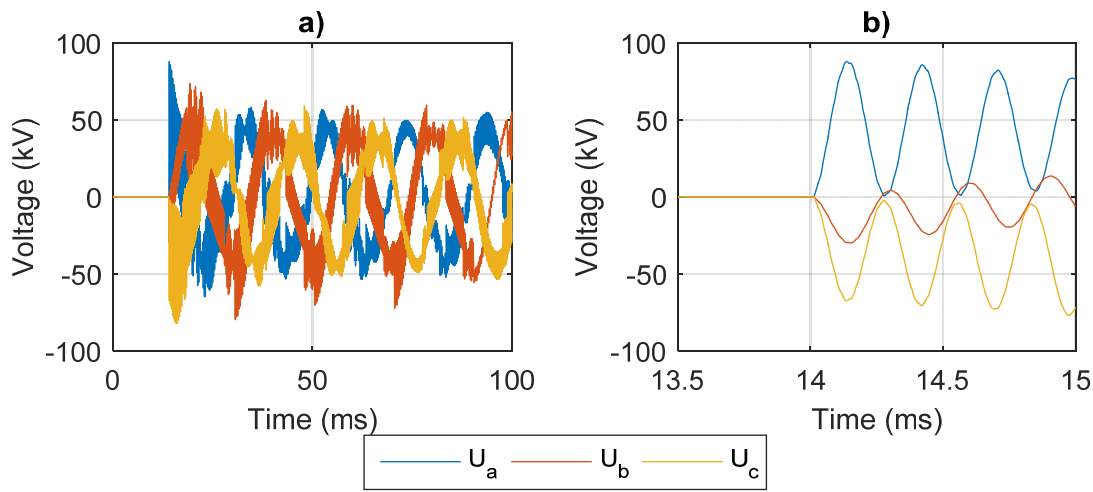

*Figure 13 Open end of 50 kV OHL for base case: a) total simulation time, b) zoomed in for initial period*

As the energization of the transformer will not take place at the maximum of the 150 kV voltage in all phases, one or two poles are closed at the switching moment around voltage minimum, giving a sinus excitation and thus a flux variation following a 1-cos response at power frequency. This limb or these limbs will go into saturation with magnetizing currents that appear also in the other phases, since there is no possibility of a zero sequence current (YΔ, with floating neutral at 150 kV-side).

### 4.2.2 Energization of the Transformer

The energization of the transformer alone is simulated. Core saturation will take place and inrush currents may appear at the 150 kV-side. Since, in this case, the 150 kV system is very strong, not much voltage deformation will be visible at the primary side, but deformation will be visible at the open 50 kV-side, as shown in Figure 15. Moreover, the transient behaviour of the transformer (Eigen-frequencies) is disregarded.
Note that the magnetizing phenomena are not influenced by the connection of the OHL.

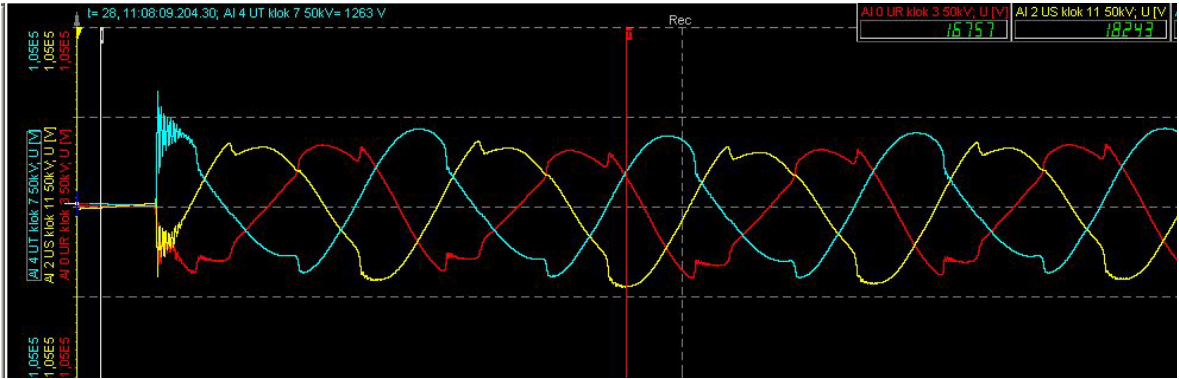

*Figure 14 Measured voltages at the open end of 50 kV OHL when energizing transformer and OHL together*

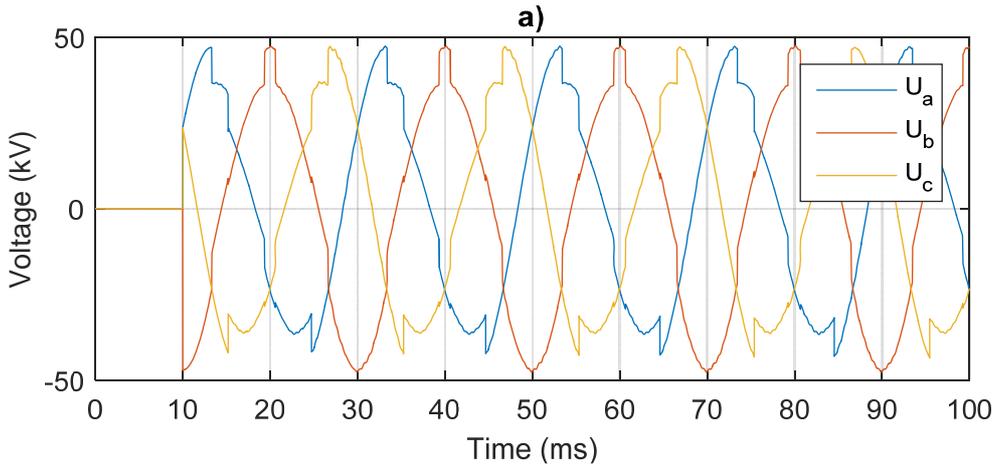

*Figure 15 Voltages at 50 kV terminals of transformer when energizing the transformer alone*

### 4.2.3 Energization of the OHL

The simulation is performed with energization of the OHL alone, i.e. the transformer is already energized in steady state. The voltages at the open end of 50 kV OHL are shown in Figure 16. Neither the magnitudes of the transient overvoltages do change, nor does the oscillation frequency. This result complies with the theoretical expectation: the same frequency for the same LC oscillation circuit.

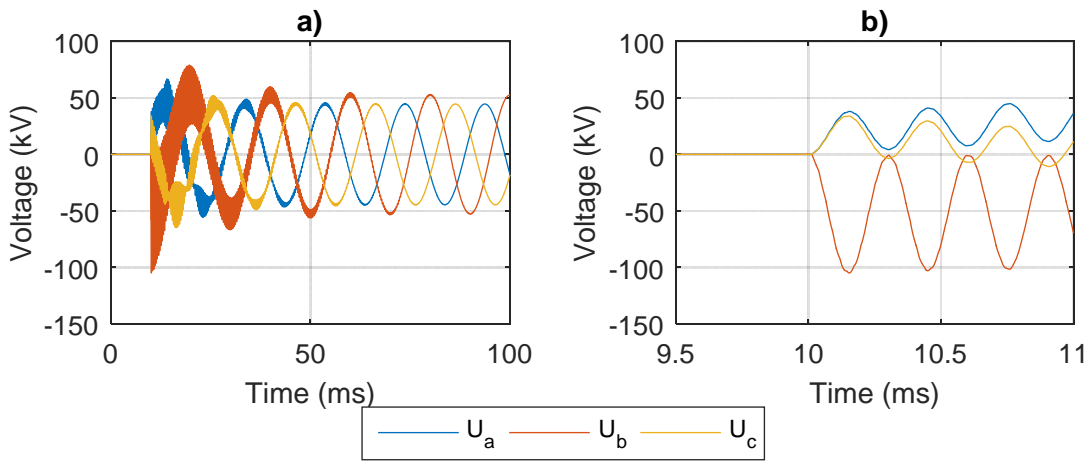

*Figure 16 Open end of 50 kV OHL when energizing the OHL alone: a) total simulation time, b) zoomed in*

### 4.2.4 Influence of Tap Positions

Simulations are performed with different tap positions of the transformer. Figure 17 shows voltages at the open end of 50 kV OHL when the tap position is at maximum transformation ratio (tap 1) and at minimum ratio (tap 21). The overvoltage is less when the tap position is at maximum ratio (tap 1), and the situation becomes worse if the tap position is at minimum ratio (tap 21).

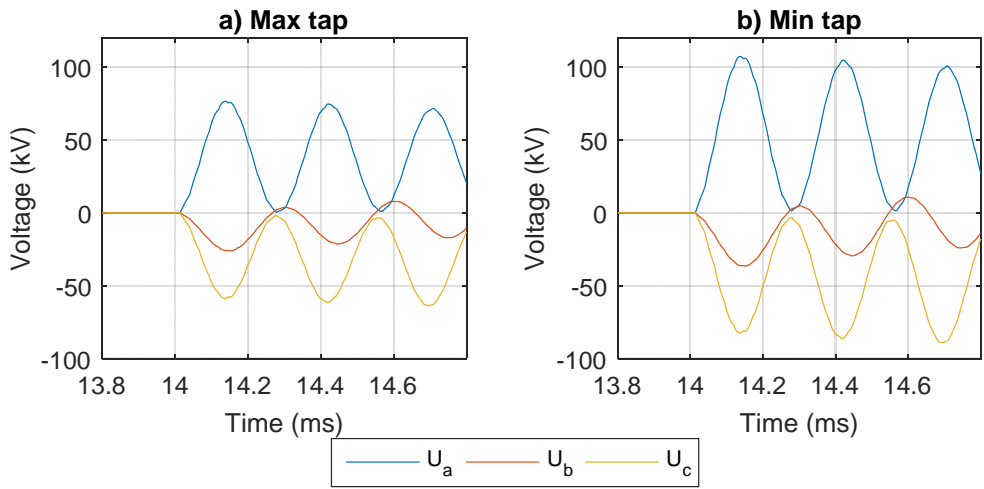

*Figure 17 Voltages at the open end of 50 kV OHL for different tap positions 1 (a) and 21 (b)*

It is clear that the transient voltage increases inversely proportional to the transformer winding ratio at 150 kV. Furthermore, the oscillation frequencies appear not to change much for different tap positions.

## 5   Conclusions

- When energizing a transformer via a long cable at 150 kV side, the voltage wave-form at the 150 kV side of the transformer shows to be independent from the tap position and the connections at the secondary side. This applies also for the voltages at other locations in the 150 kV grid. The waveform is essentially a square pattern of travelling waves, that reflect positive at the transformer (doubling its amplitude) and negative at the feeding substation. This pattern is transferred to the secondary side of the energized transformer. Initially the step function is transferred in a capacitive way, followed by the transfer function that shows the dominant Eigen- frequencies of the transformer.
- When a cable or OH-line is connected to the secondary side, the natural frequency will be dominated by the capacitance of the cable or line and the transformer's leakage inductance. With a damped 1-cos function, the voltage waveform follows the square wave-forms. The overshoot, typically for an 1-cos function, leads to higher switching overvoltages at the 50 kV side than without cables/line. In addition, travelling waves appear on the OH-line, but are hardly visible, since they cannot be built up due to the time constant of the surge impedance with the leakage inductance, even not at longer (still moderate) line lengths.
- The tap position has an influence on the voltage level at the secondary side and on the transfer function as well as some influence on the transformer leakage inductance. The tap with the lowest ratio of windings (at the 150 kV side) gives the highest switching overvoltages at the 50 kV side, but the voltages are well below the specified withstand strength of the transformer and connected equipment.
- However, in case of coinciding resonances higher switching voltages may be expected, for instance, by travelling waves that resonate with one of the transformer's Eigen-frequencies. Neither the simulations nor the service experience did show any problem in the cases investigated.
- Due to a lack of information on the complete transient functions of the transformer, it was not possible to perform an EMTP-analysis that covers the capacitive behaviour (in μs time frame), the transformer Eigen-frequencies (tens to hundreds of μs), travelling wave patterns (hundreds of μs), resonance phenomena (hundreds of μs to ms) and power frequency range in one model.